\documentclass[onecolumn]{emulateapj}

\begin{document}
\title{Constraining the Accretion Flow in Sgr A* by \\GR Dynamical and Radiative Modeling}   %%% Fill in title
\author{Roman V. Shcherbakov\altaffilmark{1}, Robert F. Penna\altaffilmark{1}}
\altaffiltext{1}{Harvard-Smithsonian Center for Astrophysics, 60 Garden Street, Cambridge, MA
02138, USA}

\begin{abstract}
We present the combination of dynamical accretion model based on 3D GRMHD simulations and general
relativistic (GR) polarized radiative transfer. We write down the formalism of and perform the GR
ray-tracing of cyclo-synchrotron radiation through the model of accretion flow in Sagittarius A*.
GR polarimetric imaging is presented as well as the results for spectrum for a probable set of
spins and orientations. Precise fitting formulae for Faraday rotation and Faraday conversion
coefficients are employed for thermal plasma. The axisymmetic flow pattern and the magnetic field
geometry correspond to averaged 3D GRMHD simulations near the black hole, whereas the analytic
model was used far from the black hole. The density scaling is found by fitting the sub-mm flux.
Spin $a=0.7$ and inclination angle $\theta=0.6$ produce the best fit to sub-mm flux and linear
polarization fraction.
\end{abstract}
\section{GR Polarized Radiative Transfer and Dynamics}
We come up with the formalism in several stages. First, we write down the standard propagation
equations of Stokes parameters $I,$ $Q,$ $U,$ $V$ in the uniform thermal plasma
\citep{melrose_dispersive} in a locally flat co-moving reference frame with synchrotron
emissivities/absorptivities from \citet{melrose_emis}. We take Faraday rotation/conversion
coefficients from \citet{shcher_farad}, as the other published derivation of Faraday conversion
coefficients \citep{huangnew} is a very crude approximation. Second, we parallel propagate the
basis vectors along the null geodesic from the observer's plane to account for GR rotation of the
basis. Third, with proper gauges on wave vector potential we write down the covariant equations of
polarized radiative transfer. Following \citet{huangnew}, we assume that the matrix of
absorptivities and propagation coefficients generalizes in the polarized transfer analogously to
the unpolarized case.

The dynamical model used in the transfer starts with adiabatic 3D GRMHD simulations of thick
accretion flow onto the Kerr black hole (BH) with spins $a=0,$ $0.7,$ $0.9,$ $0.98.$ We average the
flow velocity, magnetic field, RMS magnetic field, gas density and pressure for the quasi-steady
period of the developed accretion and separate the electron temperature $T_e$ from the proton
temperature $T_p$ by applying the heating prescription from \citet{sharma_heating}. The dynamical
model is smoothly extended to large radii $r>24M$ to take into account the Faraday rotation effect
at large distances from the BH.

%\plotone{}
 \begin{figure}[!ht]\label{fig}
\plottwo{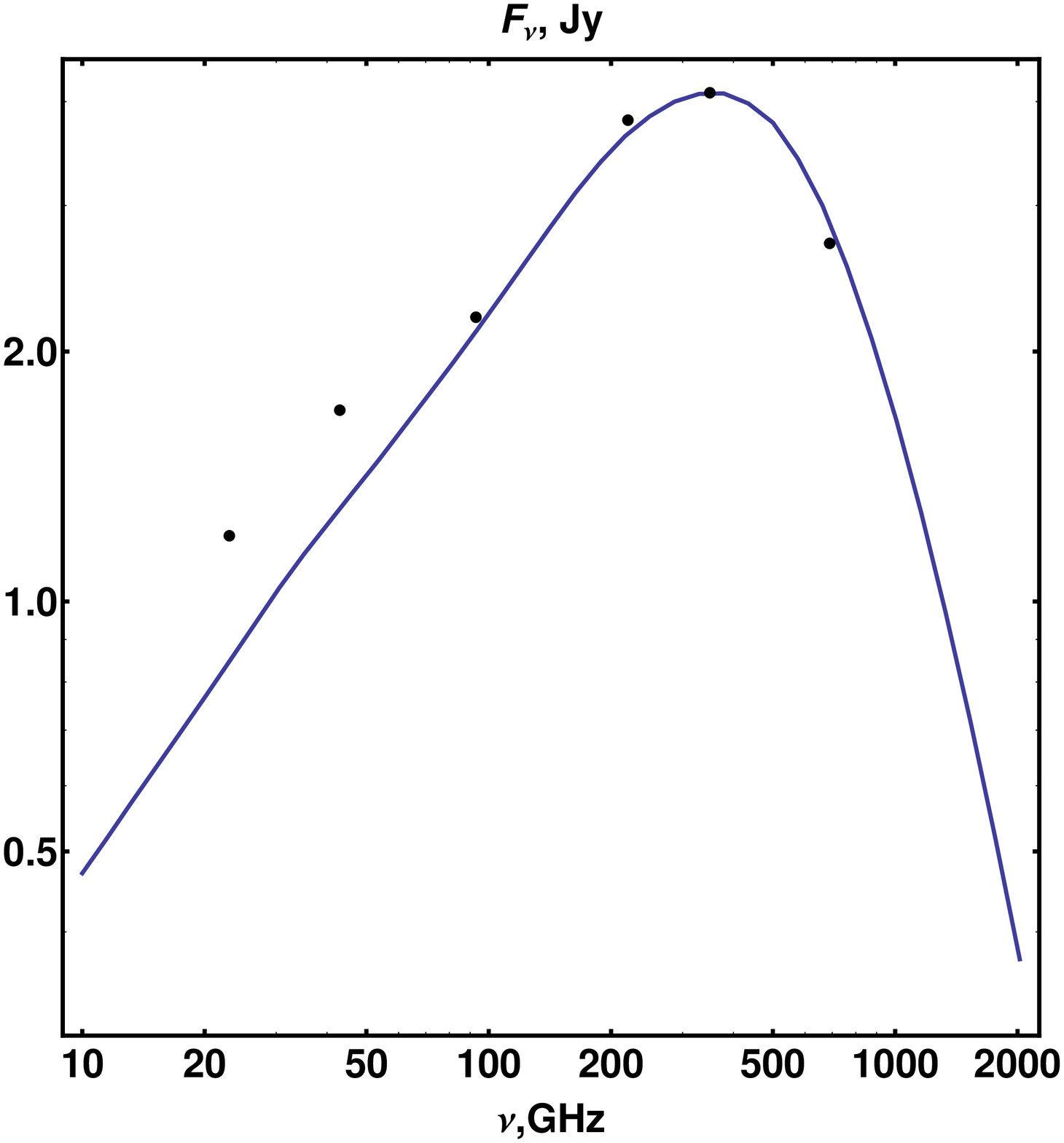}{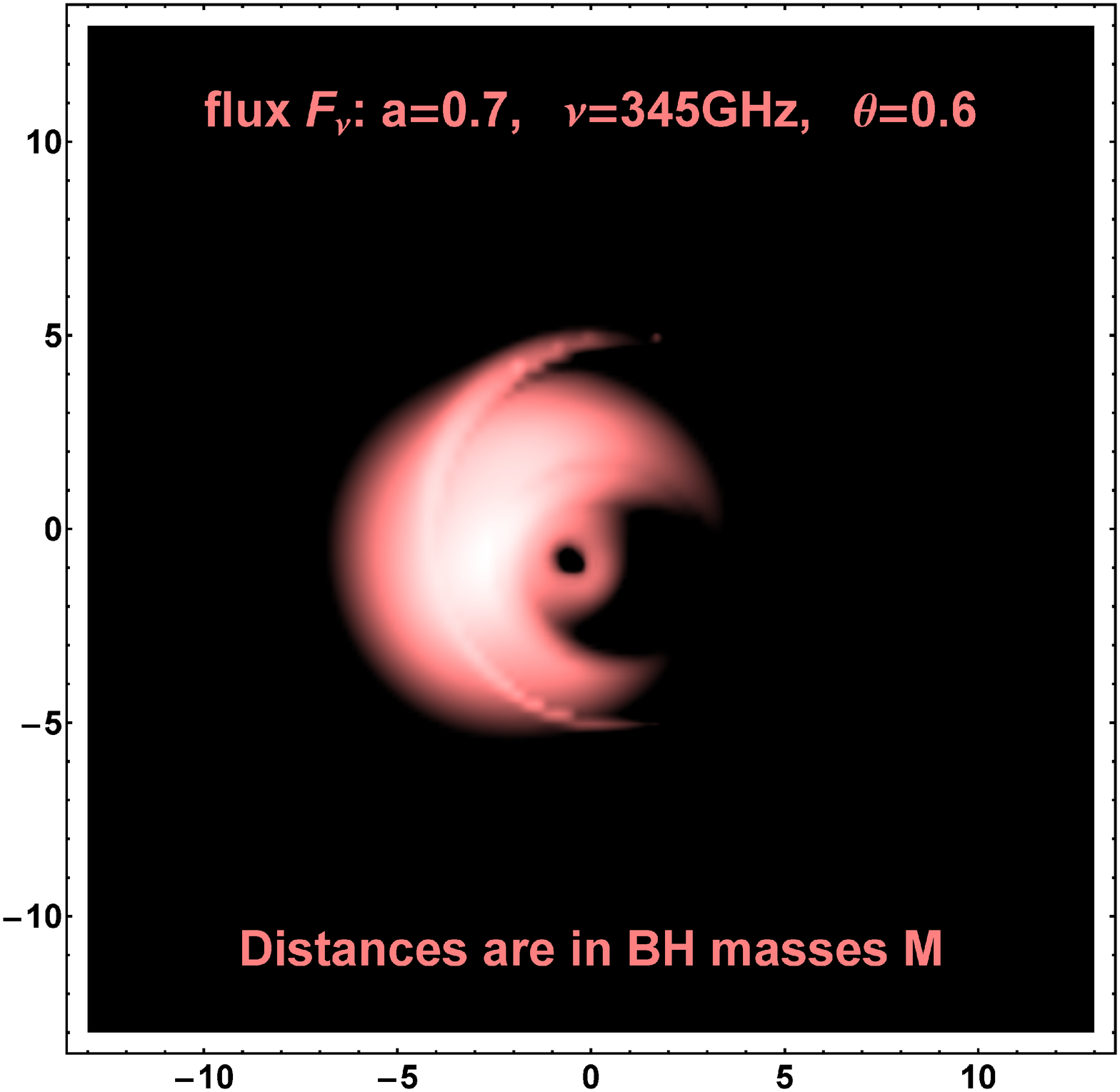} \caption{Specific flux $F_\nu$ in
comparison to observations \citep{yuan_data,marrone} (dots) on the left panel. Image of specific
flux $F_\nu$ in logarithmic scale with contrast $8$ at $\nu=345$~GHz on the right panel. In both
calculations spin $a=0.7,$ inclination angle $\theta=0.6.$}
\end{figure}

\section{Results}
We perform the GR polarized radiative transfer for each spin for a set of inclination angles
$\theta$ and compare the specific fluxes, linear polarization fractions (LP), and circular
polarization fractions (CP) to observations. We find, that the extreme spins $a=0.9,$ $a=0.98$ do
not fit all the observations well. They require lower density $n<2\cdot10^6$~cm${}^{-3}$ near BH to
fit the flux at $220$~GHz, but Faraday depolarization fails at these densities leading to high LP
at $86$~GHz compared to the observed value \citep{macquart}. The preferred value of the inclination
angle $\theta=0.6$ is coincident with that in \citet{huangnew}. The spin value $a=0.7$ gives the
best fit (see Fig.~\ref{fig}), though spin $a=0$ produces good fits as well. Imaging produces some
unexpected results. As our simulation appears to be substantially sub-Keplerian and have
significant thermal support, the Innermost Stable Circular Orbit (ISCO) loses its importance and
the black hole shadow is not always seen.

\acknowledgements Supported by NASA ESSF to RVS.

\end{document}